\documentclass[aps,showpacs,twocolumn]{revtex4}
\usepackage{epsfig}

\begin{document}

\title{X(1835), X(2120) and X(2370) in flux tube models}

\author{Chengrong Deng$^1$, Jialun Ping$^2$\footnote{Corresponding author: J. Ping (jlping@njnu.edu.cn)},
Youchang Yang$^3$, Fan Wang$^4$}

\affiliation{$^1$School of Mathematics and Physics, Chongqing
Jiaotong University, Chongqing 400074, P.R. China}

\affiliation{$^2$Department of Physics, Nanjing Normal University,
Nanjing 210097, P.R. China}

\affiliation{$^3$Department of Physics, Zunyi Normal College,
Zunyi 563002, P.R. China}

\affiliation{$^4$Department of Physics, Nanjing University,
Nanjing 210093, P.R. China}

\begin{abstract}
Nonstrange hexaquark state $q^3\bar{q}^3$ spectrum is
systematically studied by using the Gaussian expansion method in
flux tube models with a six-body confinement potential. All the
model parameters are fixed by baryon properties, so the calculation of
hexaquark state $q^3\bar{q}^3$ is parameter-free. It is found
that some ground states of $q^3\bar{q}^3$ are stable against
disintegrating into a baryon and an anti-baryon. The main components
of $X(1835)$ and $X(2370)$, which are observed in the radiative
decay of $J/\psi$ by BES collaboration,
can be described as compact hexaquark states $N_8\bar{N}_8$ and
$\Delta_8\bar{\Delta}_8$ with quantum numbers
$I^GJ^{PC}=0^+0^{-+}$, respectively. These bound states should be
color confinement resonances with three-dimensional configurations
similar to rugby ball, however, $X(2120)$ can not be accommodated
in this model approach.
\end{abstract}


\maketitle

\section{introduction}
In 2003, X(1860) was observed in the $p\bar{p}$ invariant mass
spectrum in the radiative decay $J/\psi\rightarrow\gamma
p\bar{p}$ by BES collaboration, the mass and the width are
$M=1859^{+3+5}_{-10-20}$ MeV
and $\Gamma<$30 MeV, respectively~\cite{bes0}. In 2005, $X(1835)$
was first observed in
$J/\psi\rightarrow\gamma\pi^+\pi^-\eta^{\prime}$ decays with a
statistical significance of 7.7 $\sigma$ by BES-II~\cite{bes2},
the parameters of $X(1835)$ are $M=1833.7\pm6.5{\pm2.7}$ MeV and
$\Gamma=67.7\pm20.3\pm7.7$ MeV. Very recently, the $X(1835)$ was
confirmed by BES-III in the radiative decay
$J/\psi\rightarrow\gamma\pi^+\pi^-\eta^{\prime}$ with mass and
width $M=1836.5\pm3.0^{+5.6}_{-2.1}$ MeV and
$\Gamma=190\pm9^{+38}_{-36}$ MeV, respectively~\cite{bes3}. The
mass is consistent with the BES-II result, while the width is
significantly larger. Meanwhile, $X(2120)$ and $X(2370)$ were also
observed in the same process, the masses and the widths are
$M_{X(2120)}=2122.4\pm6.7^{+4.7}_{-2.7}$ MeV,
$M_{X(2370)}=2376.3\pm8.7^{+3.2}_{-4.3}$ MeV,
$\Gamma_{X(2120)}=83\pm16^{+31}_{-11}$ MeV, and
$\Gamma_{X(2370)}=83\pm17^{+44}_{-6}$ MeV, respectively.

Various theoretical works were stimulated to interpret the natures
and structures of these resonances. Datta and O'Donnell described
$X(1860)$ as a zero baryon number, deuteron-like singlet $p\bar{p}$
$^0S_1$ state in a simple potential model with a
$\mathbf{\lambda}\cdot\mathbf{\lambda}$ confining
interaction~\cite{datta}. Ding and Yan discussed $X(1860)$ as a
baryonium and investigated mesonic decays of $X(1860)$ due to the
nucleon-antinucleon annihilation~\cite{Ding0}. Gao and Zhu
understood $X(1860)$ as the $p\bar{p}$ bound state with quantum
numbers $I^GJ^{PC}=0^+0^{-+}$, and demonstrated that it can not
decay into final state $\pi^+\pi^-$, $2\pi^0$, $\bar{K}K$ and
$3\pi$~\cite{Gao}. Kochelev and Min explained $X(1835)$ as the
lowest pseudoscalar glueball state due to the instanton mechanism of
partial $U(1)_A$ symmetry restoration~\cite{Kochelev}. He {\em et
al.} studied $X(1835)$ using the QCD sum rule and interpreted it as
a pseudoscalar state with a large gluon content~\cite{He}. Li
investigated $X(1835)$ as a $0^{-+}$ pseudoscalar glueball using an
effective Lagrangian approach~\cite{Li}. Ding {\em et al.} treated
$X(1835)$ as a baryonium with a sizable gluon content~\cite{Ding1}.
Liu proposed that X(1835) contained a baryonium component from the
large-$N_c$ QCD point of view~\cite{liuchun}. Dedonder {\em et al.}
studied $X(1835)$ in the conventional $N\bar{N}$ potential model and
suggested that it could be a broad and weakly bound state
$N\bar{N}_s(1870)$ in the $^{1}S_0$ wave. Huang and Zhu treated
X(1835) as the second radial excitation of $\eta^{\prime}(958)$ and
discussed the strong decay behavior by the effective Lagrangian
approach~\cite{Huang}. Li and Ma studied several two-body strong
decays of X(1835) associated with $\eta(1760)$ by the
quark-pair-creation model, where $X(1835)$ is assigned as the
$n^{2s+1}L_J = 3^1S_0$ $q\bar{q}$ state. Entem and Fern\'{a}ndez
derived a $N\bar{N}$ interaction from a constituent quark model
constrained by the $NN$ sector to investigate the possible baryonium
resonant state $X(1835)$~\cite{entem}. Yu {\em et al\,}'s study
indicated that: (1) X(1835) could be the second radial excitation of
$\eta^{\prime}(958)$; (2) X(2120) and (2370) can be explained as the
third and fourth radial excitations of
$\eta(548)/\eta^{\prime}(958)$~\cite{Yu}.

Quantum Chromodynamics (QCD) is widely accepted as the fundamental
theory to describe the hadron and the strong interaction and has
verified in high momentum transfer process. In the low energy
region, such as hadron spectroscopy and hadron-hadron interaction
study, the \emph{ab initio} calculation directly from QCD becomes
very difficult due to the complication of nonperturbative nature.
Recently, lattice QCD (LQCD) and nonperturbative QCD method have
made impressive progresses on hadron properties, even on
hadron-hadron interactions~\cite{Maris,Ishii1,Kanada,Inoue,Ishii2}.
However, QCD-inspired constituent quark model (CQM) is still an
useful tool in obtaining physical insight for these complicated
strong interaction systems. CQM can offer the most complete
description of hadron properties and is probably the most
successful phenomenological model of hadron
structure~\cite{godfrey}. In traditional CQM, a two-body
interaction proportional to the color charges
$\mathbf{\lambda}_i\cdot\mathbf{\lambda}_j$ and $r_{ij}^n$, where
$n=1$ or 2 and $r_{ij}$ is the distance between two quarks, was
introduced to phenomenologically describe quark confinement
interaction. The model can automatically prevent overall color
singlet multiquark states disintegrating into several color sub-systems
by means of color confinement with an appropriate
$SU_C(3)$ Casimir constant~\cite{Weinstein0}. The model also
allows a multiquark system disintegrating into
color-singlet clusters, and it leads to interacting potentials
within mesonlike $q\bar{q}$ and baryonlike $qqq$ subsystems in accord
with the empirically known potentials~\cite{Weinstein0}. However,
the model is known to be flawed phenomenologically because it
leads to power law van der Waals forces between color-singlet
hadrons~\cite{Feinberg,Weinstein1,GL81,oka1,oka2}. It is also
flawed theoretically in that it is very implausible that the
long-range static multibody potential is just a sum of the
two-body ones~\cite{Weinstein0}. The problems are related to
the fact that this model does not respect local color gauge
invariance~\cite{Lipkin,Greenberg0,Greenberg1,Robson}. Robson
proposed to use many-body confinement potentials for meson-meson and
baryon-baryon systems~\cite{Robson}, which contains the essential features of
the solution which emerges from the flux model based on the
strong coupling limit of LQCD Hamiltonian and on the explicit
local color gauge invariance~\cite{flux}.

QCD does not deny the existence of multiquark states although
experimental candidates have not been confirmed up to now. The
structures of multiquark systems and hadron-hadron interactions
are abundant~\cite{ww,ping,deng}, which have important information
that is absent in ordinary hadrons, such as $qq\bar{q}$ and
$q\bar{q}\bar{q}$ interactions~\cite{Dmitrasinovic}.
Recently, LQCD calculations on mesons, baryons, tetraquark and
pentaquark states reveal flux-tube or string like
structure~\cite{lattice1,lattice2,lattice3,lattice4}. The
confinement of multiquark states are multibody interactions and
can be simulated by a potential which proportional to the minimum
of the total length of strings which connect the quarks to form a
multiquark system. A naive flux-tube or string model basing on this
picture has been constructed~\cite{ww,ping,deng}. It takes into
account of multibody confinement with harmonic interaction
approximation, i.e., where the length of string is replaced by the
square of the length to simplify the numerical calculation. There
are two arguments to support this approximation: One is that the spatial
variations in separation of the quarks (lengths of the string) in
different hadrons do not differ significantly, so the difference
between the linear and quadratic forms is small and can be absorbed in
the adjustable parameter, the stiffness. The calculations on
nucleon-nucleon interactions support the argument~\cite{ping,npaping,sala}.
The second is that we are
using a nonrelativistic description of the dynamics and, as was
shown long ago~\cite{Goldman}, an interaction energy that varies
linearly with separation between fermions in a relativistic, first
order differential dynamics has a wide region in which a harmonic
approximation is valid for the second order (Feynman-Gell-Mann)
reduction of the equations of motion.

The flux tube model has been applied to the study of exotic
mesons~\cite{deng}. The results suggest that the multibody
confinement should be employed in the quark model study of
multiquark systems instead of the additive two-body confinement.
The flux tube model with four-body confinement potential also
described light scalar meson spectrum well in the framework of a
tetraquark picture~\cite{lightmeson}. This paper extends the model
to hexaquark $q^3\bar{q}^3$ system, to investigate systematically
the non-strange baryonium states with six-body confinement
potential. The numerical results are obtained by Gaussian
Expansion Method (GEM)~\cite{GEM}. The paper is organized as
follows: the model Hamiltonian and wavefunction for 3-quark
system are presented in Sec. II. The six-body confinement
potential and the wavefunction of a hexaquark system are
introduced in Sec. III. Sect. IV presents the numerical results
and discussions. A brief summary is given in the last section.

\section{quark models and model parameters}

The non-relativistic quark model was formulated under the
assumption that the hadrons are color singlet non-relativistic
bound states of constituent quarks with phenomenological effective
masses and interactions.

\subsection{Isgur-Karl model}
Isgur-Karl  model incorporating effective one gluon exchange (OGE)
and confinement potentials successfully describe the properties of
baryon spectrum~\cite{isgurkarl1,isgurkarl2,isgurkarl3}. The model
Hamiltonian used for baryons takes the form
\begin{eqnarray}
H & = & \sum_{i=1}^3 \left( m_i+\frac{\mathbf{p}_i^2}{2m_i}
\right)-T_{CM}+\sum_{i>j}^3V_{ij}^G +V^C,\\
V_{ij}^G & = & {\frac{1}{4}}\alpha _{s}\mathbf{\lambda}
_{i}\cdot \mathbf{\lambda}_{j}\left[{\frac{1}{r_{ij}}}-{\frac{\pi}{2}}
\delta(\mathbf{r}_{ij})
\right. \nonumber \\
&\times & \left. \left(
{\frac{1}{m_i^2}}+{\frac{1}{m_j^2}}+{\frac{4}{3m_im_j}}\mathbf{\sigma}_{i}\cdot
\mathbf{\sigma}_{j} \right) \right],\\
V^C&=&
\frac{K}{3}\left[(\mathbf{r}_1-\mathbf{r}_2)^2+(\mathbf{r}_1-\mathbf{r}_3)^2
+(\mathbf{r}_2-\mathbf{r}_3)^2\right],
\end{eqnarray}
the confinement potential $V^C$ can also be written as
\begin{eqnarray}
V^C&=&
K\left[\left(\frac{\mathbf{r}_1-\mathbf{r}_2}{\sqrt{2}}\right)^2
+\left(\frac{\mathbf{r}_1+\mathbf{r}_2-2\mathbf{r}_3}{\sqrt{6}}\right)^2\right].
\end{eqnarray}
When the model is extended to study multiquark
states~\cite{isgurkarl4}, the confinement can be equivalently
expressed as
\begin{eqnarray}
V^C&=&\sum_{i>j}^n-a_c\mathbf{\lambda}_{i}\cdot\mathbf{\lambda}_{j}r^2_{ij}.
\end{eqnarray}
Where $T_{CM}$ is the center-of-mass kinetic energy,
$\mathbf{r}_i$, $m_i$ and $\mathbf{p}_i$ are the position, mass
and momentum of the $i$-th quark, $\mathbf{\lambda}$ and
$\mathbf{\sigma}$ are the $SU(3)$ Gell-man and $SU(2)$ Pauli
matrices, respectively. Note that
$\mathbf{\lambda}\rightarrow-\mathbf{\lambda}^{*}$ for anti-quark.
All other symbols have their usual meaning. An effective
scale-dependent strong coupling constant~\cite{slamanca} is used
here
\begin{equation}
\alpha_s(\mu)=\frac{\alpha_0}{\ln\left[\frac{\mu^{2}+\mu_0^2}{\Lambda_0^2}\right]}
\end{equation}
where $\mu$ is the reduced mass of two interactional quarks, and
$\alpha_0,~\mu_0$ and $\Lambda_0$ are determined below.
The $\delta$ function, arising as a consequence of the
non-relativistic reduction of the one-gluon exchange diagram
between point-like particles, has to be regularized in order to
perform numerical calculations. It reads~\cite{bhad}
\begin{equation}
\delta(r_{ij})=\frac{1}{\beta^3\pi^{\frac{3}{2}}}e^{-r^2_{ij}/\beta^2}
\end{equation}
where $\beta$ is the model parameter which is determined by
fitting the experiment data.

\subsection{Chiral quark model}
The $SU(2)\times SU(2)$ chiral quark model described $NN$ phase shifts
and the properties of deuteron quite well~\cite{Obukhovsky,Valcarce1,Valcarce2}.
Subsequently, the $SU(3)\times SU(3)$ chiral quark model where constituent quarks
interact only through pseudoscalar Goldstone bosons exchange (GBE)
was developed to describe the baryon spectra~\cite{GlozmanRiska}.
The model including both OGE and GBE was
successfully applied to the $NN$ and nucleon-hyperon
interactions~\cite{Fujiwara1,Fujiwara2,Fujiwara3}. The Goldstone
bosons exchange potentials can be expressed as,
\begin{eqnarray}
 V_{ij}^B & = &
 v^{\pi}_{ij} \sum_{a=1}^3 \mathbf{F}_i^a \mathbf{F}_j^a
+v^{K}_{ij} \sum_{a=4}^7 \mathbf{F}_i^a \mathbf{F}_j^a
\nonumber \\
& & +v^{\eta}_{ij} (\mathbf{F}^8_i \mathbf{F}^8_j\cos \theta_P
 -\mathbf{F}^0_i \mathbf{F}^0_j\sin \theta_P),  \\
v^{\chi}_{ij} & = &
\frac{g^2_{ch}}{4\pi}\frac{m^3_{\chi}}{12m_im_j}
\frac{\Lambda^{2}_{\chi}}{\Lambda^{2}_{\chi}-m_{\chi}^2}
 \boldmath{\mbox{$\sigma$}}_i\cdot
 \boldmath{\mbox{$\sigma$}}_j \nonumber \\
 & & \left[ Y(m_\chi r_{ij})-
\frac{\Lambda^{3}_{\chi}}{m_{\chi}^3}Y(\Lambda_{\chi} r_{ij})
\right],\chi=\pi,K,\eta,\\
V^{\sigma}_{ij} & = &-\frac{g^2_{ch}}{4\pi}
\frac{\Lambda^{2}_{\sigma}}{\Lambda^{2}_{\sigma}-m_{\sigma}^2}
 m_{\sigma}\left[ Y(m_\sigma r_{ij})-
\frac{\Lambda_{\sigma}}{m_{\sigma}}Y(\Lambda_{\sigma}r_{ij})
\right].\nonumber
\end{eqnarray}
where $Y(x)$ is the standard Yukawa functions defined by
$Y(x)=\frac{e^{-x}}{x}$ and $\mathbf{F}^a$ is flavor $SU(3)$
Gell-mann matrices. The angle $\theta_{P}$ appears as a consequence
of considering the physical $\eta$ instead of the octet one.
$m_{\pi}$, $m_K$ and $m_{\eta}$ are the masses of the $SU(3)$
Goldstone bosons, took their experimental values. $m_{\sigma}$ is
determined through the PCAC relation $m^2_{\sigma}\sim
m^2_{\pi}+4m^2_{u,d}$~\cite{Scadron}. The chiral coupling constant
$g_{ch}$ is determined from the $\pi NN$ coupling constant through
\begin{equation}
\frac{g_{ch}^2}{4\pi}=\left(\frac{3}{5}\right)^2\frac{g_{\pi
NN}^2}{4\pi}\frac{m_{u,d}^2}{m_N^2}
\end{equation}
Here the flavor $SU(3)$ is assumed to be an exact system and only broken
by the different mass of the strange quark. The confinement and OGE
interaction terms are the same as those of Isgur-Karl model and will
not be rewritten here.

\begin{figure}[ht]
\epsfxsize=2.8in \epsfbox{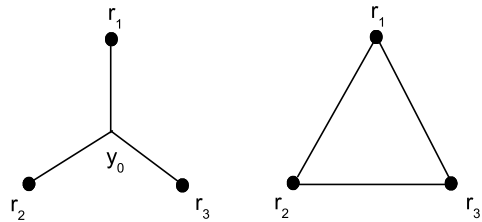} \caption{Three-body and two-body
confinement potential}
\end{figure}

\subsection{Flux-tube model}
This model assumption is inspired by the LQCD calculation. LQCD
calculations for baryons reveal flux-tube or string like
structure~\cite{lattice2,Bissey}. The simplified version, Y-shape
structure, is shown in Fig. 1, where the $\mathbf{r}_i$ represents
the spatial position of the $i$-th quark denoted by a black dot and
$\mathbf{y}_0$ denotes the junction where three color flux tubes
meet. The confinement is proportional to the minimum of the sum of
the square of the length of three flux tubes. In the flux tube model
with quadratic potential, the three-body confinement can be written
as
\begin{eqnarray}
V^C=K\left[(\mathbf{r}_1-\mathbf{y}_0)^2+(\mathbf{r}_2-\mathbf{y}_0)^2
+(\mathbf{r}_3-\mathbf{y}_0)^2\right]
\end{eqnarray}
For the confinement potential $V^C$, the position of the junction
$\mathbf{y}_0$ can be fixed by minimizing the energy of the
system, then we get
\begin{equation}
\mathbf{y}_0=\frac{\mathbf{r}_1+\mathbf{r}_2+\mathbf{r}_3}{3}
\end{equation}
Therefore, the minimum of the confinement potential for baryons
$V_{min}^C$ has the following forms
\begin{equation}
V_{min}^{C}=
K\left[\left(\frac{\mathbf{r}_1-\mathbf{r}_2}{\sqrt{2}}\right)^2
+\left(\frac{2\mathbf{r}_3-\mathbf{r}_1-\mathbf{r}_2}{\sqrt{6}}\right)^2\right].
\end{equation}
The other parts of Hamiltonian of flux tube model are the same as
Isgur-Karl model (denoted as Model I hereafter) or chiral quark model
(Model II). It should be noted that for a baryon the three-body
quadratic confinement potential is exactly equivalent
to the sum of two-body one, $\Delta$-shape in Fig. 1 (although it is not
exactly the case for the linear confinement potential). As far as baryon
is concerned, the flux tube model is not a new model. However, when it
is applied to multiquark systems, the flux-tube confinement potential is
different from the traditional two-body confinement ((Isgur-Karl model
and chiral quark model)~\cite{ping,deng,ww}.

In this work, the tensor forces and spin-orbit forces between
quarks are omitted in three models, because of their small or
zero contributions to the ground state baryons.

\subsection{Wavefunctions and baryon spectrums}
For baryons, the color part wavefunction $\psi_c$ is
antisymmetrical because of the color singlet requirement. The
spatial wavefunction $\psi^G_{L_TM_T}(\mathbf{R}, \mathbf{r})$ is
assumed to be symmetrical because we are interested in ground
states. So the spin-flavor wavefunction $\psi_{IM_ISM_S}$, the
$SU(6)\supset SU_s(2)\times SU_f(3)$ symmetry is used here, is
symmetrical under the exchange of two identical particles. The
total antisymmetrical wavefunction can be described as,
\begin{equation}
\Phi_{IM_IJM_J}(\mathbf{R}, \mathbf{r})=\psi_c
\left[\psi^G_{L_TM_T}(\mathbf{R},
\mathbf{r})\psi_{IM_ISM_S}\right]_{IM_IJM_J}.
\end{equation}
$[\cdots]_{IM_IJM_J}$ means coupling the spin $S$ and total
orbital angular momentum $L_T$ with Clebsch-Gordan coefficients.

We define Jacobi coordinates $\mathbf{r}_{ij}$ and
$\mathbf{R}_k$ for the cyclic permutations of $(1,2,3)$,
\begin{eqnarray}
\mathbf{r}_{ij}=\mathbf{r}_i-\mathbf{r}_j,\ {} \ {}
\mathbf{R}_k=\mathbf{r}_k-\frac{m_i\mathbf{r}_i+m_j\mathbf{r}_j}{m_i+m_j}.
\end{eqnarray}
Then, the spatial symmetrical wavefunctions can be expressed as,
\begin{eqnarray}
\Psi_{L_TM_T}(\mathbf{R}, \mathbf{r})=
{\sum_{i,j,k=1}^{3}}\left[\phi_{lm}(\mathbf{r}_{ij})
\phi_{LM}(\mathbf{R}_k)\right]_{L_TM_T},
\end{eqnarray}
$\phi_{lm}(\mathbf{r}_{ij})$ and $~\phi_{LM}(\mathbf{R}_k)$ are
the superpositions of Gaussian basis functions with different
sizes,
\begin{eqnarray}
\phi_{lm}(\mathbf{r}_{ij})&=&\sum_{n=1}^{n_{max}}c_nN_{nl}r_{ij}^{l}
e^{-\nu_nr_{ij}^2}Y_{lm}(\hat{\mathbf{r}_{ij}}), \\
\psi_{LM}(\mathbf{R}_k)&=&\sum_{N=1}^{N_{max}}c_NN_{NL}R_k^{L}
e^{-\nu_{N}R_k^2}Y_{LM}(\hat{\mathbf{R}_k}),
\end{eqnarray}
where $N_{nl}$ and $N_{NL}$ are normalization constants. Gaussian
size parameters $\nu_{n}$ and $\nu_{N}$ are taken as geometric
progression,
\begin{eqnarray}
r_n=r_1a^{n-1},&\nu_{n}=\frac{1}{r^2_n},&
a=\left(\frac{r_{n_{max}}}{r_1}\right)^{\frac{1}{n_{max}-1}},\\
R_N=R_1A^{N-1},&\nu_{N}=\frac{1}{R^2_N},&
A=\left(\frac{R_{N_{max}}}{R_1}\right)^{\frac{1}{N_{max}-1}}.
\end{eqnarray}
The numbers $n$ and $l$ ($N$ and $L$) specify the radial and angular
momenta excitations with respect to the Jacobi coordinates
$\mathbf{r}$ $(\mathbf{R})$, respectively. The angular momenta $l$
and $L$ are coupled to the total orbit angular momentum $L_T$. In
the present work all three angular momenta are assumed to be zero.

Using above Hamiltonian and wavefunctions, the light baryon spectra
and the corresponding model parameters can be obtained
by solving the three-body Schr\"{o}dinger equation
\begin{eqnarray}
(H_3-E)\Phi_{IM_IJM_J}(\mathbf{R},\mathbf{r})=0
\end{eqnarray}
with Rayleigh-Ritz variational principle. The converged results,
which are shown in Table I, are arrived by setting $r_1=R_1=0.3$
fm, $r_{n_{max}}=R_{n_{max}}=2.0$ fm and $n_{max}=N_{max}=5$. It
can be seen from Table I that Isgur-Karl model and chiral quark
model give similar numerical results, which can describe well the
light baryon spectrum.
\begin{table}[ht]
\caption{Baryon spectra (unit: MeV).}
\begin{tabular}{cc|ccccccccccc}\hline\hline
State&
&&N&$\Lambda$&$\Sigma$&$\Xi$&$\Delta$&$\Sigma^*$&$\Xi^*$&$\Omega$
 \\\hline
 Isgur-Karl &&& 939 & 1022 &1196 & 1307 & 1232 & 1397 & 1542 & 1673\\
 Chiral     &&& 939 & 1048 &1249 & 1375 & 1232 & 1391 & 1536 & 1670\\
 Expreiment &&& 939 & 1116 &1195 & 1315 & 1232 & 1384 & 1533 & 1672\\
\hline\hline
\end{tabular}
\end{table}
\begin{table}[ht]
\caption{Model parameters.}
\begin{tabular}{c|c|c|cccccccc}\hline\hline
Classification&~ Parameters ~&~~Isgur-Karl~~&~~Chiral~~  \\
  &~  ~&~~(Model I)~~&~~(Model II)~~  \\
\hline
              &~ $m_{ud}$ (MeV)                ~&~~  313     ~~&~~360      \\
              &~ $m_{s}$ (MeV)                 ~&~~  585     ~~&~~560      \\
Re-adjusted   &~ K~(MeV fm$^{-2}$)             ~&~~  336     ~~&~~224      \\
              &~ $\beta$ (fm)                  ~&~~  0.32    ~~&~~0.08     \\
              &~ $\alpha_0$                    ~&~~  6.82    ~~&~~5.21     \\
\hline
              &~ $\Lambda_0$ (fm$^{-1}$)       ~&~~  0.187   ~~&~~0.187    \\
              &~ $\mu_0$  (fm$^{-1}$)          ~&~~  0.113   ~~&~~0.113    \\
              &~ $\Lambda_{\pi}$(fm$^{-1}$)    ~&~~  ---     ~~&~~4.2      \\
              &~ $\Lambda_{\sigma}$(fm$^{-1}$) ~&~~  ---     ~~&~~4.2      \\
              &~ $\Lambda_{K}$     (fm$^{-1}$) ~&~~  ---     ~~&~~5.2      \\
Fixed         &~ $\Lambda_{\eta}$  (fm$^{-1}$) ~&~~  ---     ~~&~~5.2      \\
              &~ $m_{\pi}$         (fm$^{-1}$) ~&~~  ---     ~~&~~0.70     \\
              &~ $m_{K}$           (fm$^{-1}$) ~&~~  ---     ~~&~~2.51     \\
              &~ $m_{\eta}$        (fm$^{-1}$) ~&~~  ---     ~~&~~2.77     \\
              &~ $m_{\sigma}$      (fm$^{-1}$) ~&~~  ---     ~~&~~3.72     \\
              &~ $\theta_{P}$      ($^{o})$    ~&~~  ---     ~~&~~-15      \\
              &~ $g_{ch}^2/{4\pi}$             ~&~~  ---     ~~&~~0.54     \\
\hline\hline
\end{tabular}
\end{table}
The fitting parameters in Isgur-Karl model and chiral quark model
are listed in Table II, in which five parameters are re-adjusted
to fit the light baryon spectrum. Other
parameters $\Lambda_{\pi}$, $\Lambda_{\sigma}$, $\Lambda_{\eta}$,
$\Lambda_{K}$, $\theta_P$, $\Lambda_0$ and $\mu_0$, which are fixed
by fitting the meson spectra, are taken from Ref.~\cite{slamanca}.

\section{six-body confinement potentials in the flux tube model}

In the flux tube model it is assumed that the color-electric flux is
confined to narrow, string-like tubes joining quarks. A flux tube
starts from every quark and ends at an anti-quark or a Y-shaped
junction, where three flux tubes annihilate or are
created~\cite{flux}. In general, a state with $N+1$-particles can
be generated by replacing a quark or an anti-quark in an
$N$-particles state by a Y-shaped junction and two quarks or two
anti-quarks.

The $q^3\bar{q}^3$ systems have been studied in the usual
constituent quark model including a two-body confinement potential
proportional to a color factor, no bound state is found for
non-strange system~\cite{entem,MPLA26}. Vijande {\em et al} recently
studied the stability of hexaquark states ($q^6$ and $q^3\bar{q}^3$)
in the string confinement and found that the ground states of
$Q^3\bar{q}^3$ are stable against disintegrating into two color
singlet baryons~\cite{hexaquarks}. For $q^3\bar{q}^3$ system, it can
be consisted of a color singlet baryon and a color singlet
anti-baryon as in the usual hadron degree of freedom description,
but also of a color octet baryon and a color octet anti-baryon
coupled to an overall color singlet six quark state. The former is
named as hadronic molecule state, the latter is called hidden color
channel and because of color confinement, the hidden color channel
exists in the two-cluster overlap region only. These two structures
are shown in Figs. 2 and 3, respectively. In general, a hexaquark
system $q^3\bar{q}^3$ should be a mixture of these two components.
These two structures for $q^3\bar{q}^3$ system are considered in the
present work.

In Figs. 2 and 3, $\mathbf{r}_{\alpha}$ represents the position
coordinate of the quark $q_{\alpha}$ (antiquark $\bar{q}_{\alpha}$)
which is denoted by a solid (hollow) dot, where ${\alpha}=i,...,n$,
$(i,j,k)$ and $(l,m,n)$ are cyclic indexes for (1,2,3) and (4,5,6),
respectively. $\mathbf{y}_{\beta}$ represents a junction, where
$\beta=1,...,4$. A thin line connecting a quark and a junction
represents a fundamental string, {\em i.e.}, a color triplet, a
thick line connecting two junctions is for color sextet, octet or
others, namely a compound string.
\begin{figure}[ht]
\epsfxsize=2.8in \epsfbox{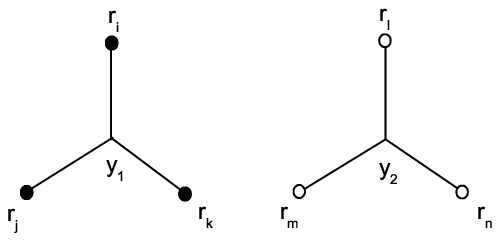} \caption{Hadronic molecule
structure.} \epsfxsize=2.5in \epsfbox{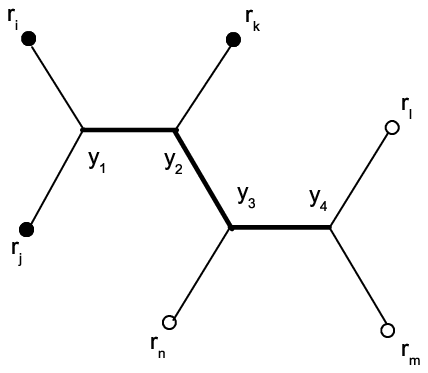} \caption{Hidden
color flux tube structure.}
\end{figure}

Within the flux tube model, the confinement potential for a
hadronic molecule state can be written as
\begin{eqnarray}
V_{min}^{CM}&=&
K\left[\left(\frac{\mathbf{r}_i-\mathbf{r}_j}{\sqrt{2}}\right)^2
+\left(\frac{2\mathbf{r}_k-\mathbf{r}_i-\mathbf{r}_j}{\sqrt{6}}\right)^2
\right. \nonumber \\
&+&
\left.\left(\frac{\mathbf{r}_l-\mathbf{r}_m}{\sqrt{2}}\right)^2
+\left(\frac{2\mathbf{r}_n-\mathbf{r}_l-\mathbf{r}_m}{\sqrt{6}}\right)^2
\right].
\end{eqnarray}
With respect to a hidden color state, the confinement potential
has the following form
\begin{eqnarray}
V^{CH}&=&K\left[(\mathbf{r}_i-\mathbf{y}_1)^2+(\mathbf{r}_j-\mathbf{y}_1)^2
+(\mathbf{r}_k-\mathbf{y}_2)^2\right. \nonumber \\
&+&
\left.(\mathbf{r}_n-\mathbf{y}_3)^2+(\mathbf{r}_l-\mathbf{y}_4)^2
+(\mathbf{r}_m-\mathbf{y}_4)^2 \right. \nonumber\\
&+&
\left.\kappa_{d_{12}}(\mathbf{y}_1-\mathbf{y}_2)^2+\kappa_{d_{23}}(\mathbf{y}_2
-\mathbf{y}_3)^2 \right.  \\
&+&\left.\kappa_{d_{34}}(\mathbf{y}_3-\mathbf{y}_4)^2
\right]. \nonumber
\end{eqnarray}
The string stiffness constant of an elementary or color triplet
string is $K$, while $K\kappa_{d_{ij}}$ is other compound string
stiffness. The compound string stiffness parameter
$\kappa_{d_{ij}}$~\cite{Bali} depends on the color dimension,
$d_{ij}$, of the string,
\begin{equation}
 \kappa_{d_{ij}}=\frac{C_{d_{ij}}}{C_3},
\end{equation}
where $C_{d_{ij}}$ is the eigenvalue of the Casimir operator
associated with the $SU(3)$ color representation $d_{ij}$ on
either end of the string, namely $C_3=\frac{4}{3}$,
$C_6=\frac{10}{3}$ and $C_8=3$. In numerical calculations, the
average $\kappa_d$ for $\kappa_{d_{ij}}$ is used for simplicity.

For given quark (antiquark) positions $\mathbf{r}_{\alpha}$, those
junction coordinates $\mathbf{y}_{\beta}$ are obtained by minimizing the
confinement potential. By introducing the following set of
canonical coordinates $\mathbf{R}_i$,
\begin{eqnarray}
\mathbf{R}_1&=&\frac{1}{\sqrt{2}}(\mathbf{r}_i-\mathbf{r}_j),
~~\mathbf{R}_2=\frac{1}{\sqrt{2}}(\mathbf{r}_l-\mathbf{r}_m)\nonumber\\
\mathbf{R}_3&=&\frac{1}{\sqrt{12}}(\mathbf{r}_i+\mathbf{r}_j
-2\mathbf{r}_k-2\mathbf{r}_n+\mathbf{r}_l+\mathbf{r}_m)\\
\mathbf{R}_4&=&\frac{1}{\sqrt{33+5\sqrt{33}}}(\mathbf{r}_i+\mathbf{r}_j
-w_1\mathbf{r}_k+w_1\mathbf{r}_n-\mathbf{r}_l-\mathbf{r}_m)\nonumber\\
\mathbf{R}_5&=&\frac{1}{\sqrt{33-5\sqrt{33}}}\left(\mathbf{r}_i+\mathbf{r}_j
+w_2\mathbf{r}_k-w_2\mathbf{r}_n-\mathbf{r}_l-\mathbf{r}_m\right)\nonumber\\
\mathbf{R}_6&=&\frac{1}{\sqrt{6}}(\mathbf{r}_i+\mathbf{r}_j+\mathbf{r}_k
+\mathbf{r}_l+\mathbf{r}_m+\mathbf{r}_n), \nonumber
\end{eqnarray}
where $w_1=\frac{\sqrt{33}+5}{2}$ and $w_2=\frac{\sqrt{33}-5}{2}$,
the minimum of the confinement potential takes the following form,
\begin{eqnarray}
V^{CH}_{min}&=&K\left[\mathbf{R}_1^2+\mathbf{R}_2^2+\frac{3\kappa_d}{2+3\kappa_d}\mathbf{R}_3^2
\right. \\&+&\left.
\frac{2\kappa_d(\kappa_d+w_3)}{2\kappa_d^2+7\kappa_d+2}\mathbf{R}_4^2
+\frac{2\kappa_d(\kappa_d+w_4)}{2\kappa_d^2+7\kappa_d+2}\mathbf{R}_5^2\right],\nonumber
\end{eqnarray}
where $w_3=\frac{7+\sqrt{33}}{4}$ and $w_4=\frac{7-\sqrt{33}}{4}$.
Clearly this confinement potential is multibody interaction rather
than the sum of two-body one in the sense that a move of a quark
may affect flux tubes connecting pattern.

When two clusters $q^3$ and $\bar{q}^3$ separate in a long distance,
a baryon and an antibaryon should be a dominant component of a hexaquark
$q^3\bar{q}^3$ system because other hidden color flux tube
structures are suppressed due to the confinement. On the other hand,
if the separation is intermediate, a hadronic molecule state may be
formed if the attractive force between a baryon and an antibaryon is strong
enough. When the two quark-clusters are close enough to be within the range of
confinement (about 1 fm), all possible flux tube structures will
appear due to the excitation and rearrangement of flux tubes. In
this case, the confinement potential of a hexaquark system
$q^3\bar{q}^3$ should be taken to be the minimum of two flux tube
structures. It reads
\begin{eqnarray}
V^C_{min} = \mbox{min} \left[ V^{CM}_{min}, V^{CH}_{min} \right].
\end{eqnarray}

\section{numerical results and discussions}
The flux tube structure specifies how the colors of quarks and
anti-quarks are coupled to form an overall color singlet.
Therefore, the model wavefunction with defined quantum numbers
$I_T$ and $J_T$ can be expressed as,
\begin{eqnarray}
\Psi_{I_T,J_T}^{q^3\bar{q}^3}=\sum
c_{\xi}\left[\left[\Phi^{q^3}_{c_1IJ}
\Phi^{\bar{q}^3}_{c_2I^{\prime}J^{\prime}}\right]_{\xi}F_{L^\prime}
(\mathbf{X})\right]_{I_TJ_T}
\end{eqnarray}
$\Phi^{q^3}_{c_1IJ}$ and
$\Phi^{\bar{q}^3}_{c_2I^{\prime}J^{\prime}}$ are cluster
wavefunctions of colorful or colorless baryon and anti-baryon,
respectively. The spatial wavefunctions are the same as those of
baryons shown before, $[\cdots]_\xi$ represents all the needed
coupling: color, isospin and spin coupling.
$F_{L^\prime}(\mathbf{X})$ is the relative orbital wavefunction
between $q^3$ and $\bar{q}^3$ clusters. All the possible channels
are taken into account in our multichannel coupling calculation, the
details can be seen in Table III.
\begin{figure}[ht]
\epsfxsize=2.8in \epsfbox{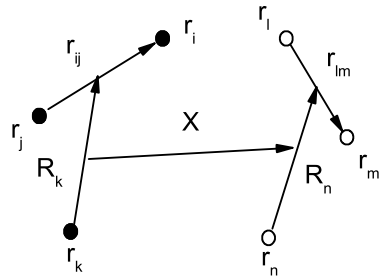} \caption{Jacobi ordinates for a
$q^3\bar{q}^3$ system.}
\end{figure}
The Jacobi coordinates for a $q^3\bar{q}^3$ system are shown in
Fig. 4, which can be expressed as
\begin{eqnarray}
\mathbf{r}_{ij}&=&\mathbf{r}_i-\mathbf{r}_j,~~
\mathbf{R}_k=\mathbf{r}_k-\frac{\mathbf{r}_i+\mathbf{r}_j}{2},\nonumber\\
\mathbf{r}_{lm}&=&\mathbf{r}_l-\mathbf{r}_m,~~
\mathbf{R}_n=\mathbf{r}_n-\frac{\mathbf{r}_l+\mathbf{r}_m}{2},\\
\mathbf{X}&=&\frac{\mathbf{r}_i+\mathbf{r}_j+\mathbf{r}_k}{3}-
\frac{\mathbf{r}_l+\mathbf{r}_m+\mathbf{r}_n}{3}. \nonumber
\end{eqnarray}
Using GEM, the relative orbital wavefunction $F_{L^\prime}(\mathbf{X})$ can
be written as,
\begin{eqnarray}
F_{L^\prime}(\mathbf{X})=\sum_{N^{\prime}=1}^{N^{\prime}_{max}}c_{N^{\prime}}
N_{N^{\prime}L^{\prime}}X^{L^{\prime}}
e^{-\nu_{N^{\prime}}X^2}Y_{L^{\prime}M^{\prime}}(\hat{\mathbf{X}})
\end{eqnarray}

Now we turn to the numerical calculations on $q^3\bar{q}^3$ systems.
In Model I and II, where a six-body confinement potential is used,
all the model parameters are fixed by fitting the ground state
baryon spectrum, no new parameter is introduced in the six-body
calculation. The eigenvalues and eigenfunctions of the
$q^3\bar{q}^3$ states can be obtained by solving the following
six-body Schr\"{o}dinger equation
\begin{eqnarray}
(H_6-E)\Psi^{q^3\bar{q}^3}_{I_TJ_T}=0
\end{eqnarray}
with Rayleigh-Ritz variational principle. The calculated results are converged
with $n_{max}$=5, $N_{max}=5$ and $N^{\prime}_{max}=5$. Minimum and maximum
ranges of the bases are 0.3 fm and 2.0 fm for coordinates $\mathbf{r}$,
$\mathbf{R}$ and $\mathbf{X}$, respectively.

\begin{table}
\caption{Binding energies of lowest $q^3\bar{q}^3$ states with all
possible quantum numbers in Model I and II. (unit: MeV)}
\begin{tabular}{c|c|c|c|ccccccc}\hline\hline
$ I^GJ^{PC}$  & Coupled channels &$E_T(B+\bar{B})$& $\Delta E_{I}$ & $\Delta E_{II}$ \\
\hline $ 0^+0^{-+}$  & $N_8^{\frac{1}{2}}\bar{N}_8^{\frac{1}{2}}$,
$\Delta_8^{\frac{1}{2}}\bar{\Delta}_8^{\frac{1}{2}}$,
$N_8^{\frac{3}{2}}\bar{N}_8^{\frac{3}{2}}$,&939+939& -44 & -34
\\ &$N\bar{N}$, $\Delta\bar{\Delta}$&&
\\  \hline
$0^-1^{--}$&$N_8^{\frac{1}{2}}\bar{N}_8^{\frac{1}{2}}$,
$N_8^{\frac{1}{2}}\bar{N}_8^{\frac{3}{2}}$,
$\Delta_8^{\frac{1}{2}}\bar{\Delta}_8^{\frac{1}{2}}$,&939+939&0&0\\
&$N_8^{\frac{3}{2}}\bar{N}_8^{\frac{1}{2}}$,
$N_8^{\frac{3}{2}}\bar{N}_8^{\frac{3}{2}}$, $N\bar{N}$,
$\Delta\bar{\Delta}$&&
\\ \hline
$0^+2^{-+}$&$N_8^{\frac{1}{2}}\bar{N}_8^{\frac{3}{2}}$,
$N_8^{\frac{3}{2}}\bar{N}_8^{\frac{1}{2}}$,&1232+1232&-269&-200\\
&$N_8^{\frac{3}{2}}\bar{N}_8^{\frac{3}{2}}$,
$\Delta\bar{\Delta}$&&
\\ \hline
$0^-3^{--}$&$N_8^{\frac{3}{2}}\bar{N}_8^{\frac{3}{2}}$,
$\Delta\bar{\Delta}$&1232+1232&0&-58
\\  \hline
$1^-0^{-+}$&$N_8^{\frac{1}{2}}\bar{N}_8^{\frac{1}{2}}$,
$N_8^{\frac{1}{2}}\bar{\Delta}_8^{\frac{1}{2}}$,
$\Delta_8^{\frac{1}{2}}\bar{N}_8^{\frac{1}{2}}$,&939+939&-44&-5\\
&$\Delta_8^{\frac{1}{2}}\bar{\Delta}_8^{\frac{1}{2}}$,
$N_8^{\frac{3}{2}}\bar{N}_8^{\frac{3}{2}}$, $N\bar{N}$,
$\Delta\bar{\Delta}$&&
\\  \hline
&$N_8^{\frac{1}{2}}\bar{N}_8^{\frac{1}{2}}$,
$N_8^{\frac{1}{2}}\bar{\Delta}_8^{\frac{1}{2}}$,
$N_8^{\frac{1}{2}}\bar{N}_8^{\frac{3}{2}}$,&&\\
$1^+1^{--}$&$\Delta_8^{\frac{1}{2}}\bar{N}_8^{\frac{1}{2}}$,
$\Delta_8^{\frac{1}{2}}\bar{\Delta}_8^{\frac{1}{2}}$,
$\Delta_8^{\frac{1}{2}}\bar{N}_8^{\frac{3}{2}}$,&939+939&0&0
\\
&$N_8^{\frac{3}{2}}\bar{N}_8^{\frac{1}{2}}$,
$N_8^{\frac{3}{2}}\bar{\Delta}_8^{\frac{1}{2}}$,
$N_8^{\frac{3}{2}}\bar{N}_8^{\frac{3}{2}}$,&&
\\
&$N\bar{N}$, $N\bar{\Delta}$, $\Delta\bar{N}$,
$\Delta\bar{\Delta}$&&
\\ \hline
&$N_8^{\frac{1}{2}}\bar{N}_8^{\frac{3}{2}}$,
$\Delta_8^{\frac{1}{2}}\bar{N}_8^{\frac{3}{2}}$,
$N_8^{\frac{3}{2}}\bar{N}_8^{\frac{1}{2}}$,&&\\
$1^-2^{-+}$& $N_8^{\frac{3}{2}}\bar{\Delta}_8^{\frac{1}{2}}$,
 $N_8^{\frac{3}{2}}\bar{N}_8^{\frac{3}{2}}$, $N\bar{\Delta}$,&939+1232&-7&-71
\\
& $\Delta\bar{N}$, $\Delta\bar{\Delta}$&&
\\  \hline
$1^+3^{--}$&$N_8^{\frac{3}{2}}\bar{N}_8^{\frac{3}{2}}$,
$\Delta\bar{\Delta}$&1232+1232&0&-44
\\ \hline
$2^+0^{-+}$&$N_8^{\frac{1}{2}}\bar{\Delta}_8^{\frac{1}{2}}$,
$\Delta_8^{\frac{1}{2}}\bar{N}_8^{\frac{1}{2}}$,&1232+1232&-88&-87
\\
& $\Delta_8^{\frac{1}{2}}\bar{\Delta}_8^{\frac{1}{2}}$,
$\Delta\bar{\Delta}$&&
\\ \hline
&$N_8^{\frac{1}{2}}\bar{\Delta}_8^{\frac{1}{2}}$,
$\Delta_8^{\frac{1}{2}}\bar{N}_8^{\frac{1}{2}}$,
$\Delta_8^{\frac{1}{2}}\bar{\Delta}_8^{\frac{1}{2}}$,&&
\\
$2^-1^{--}$& $\Delta_8^{\frac{1}{2}}\bar{N}_8^{\frac{3}{2}}$ ,
$N_8^{\frac{3}{2}}\bar{\Delta}_8^{\frac{1}{2}}$,
$N\bar{\Delta}$,&939+1232&-13&-108
\\
& $\Delta\bar{N}$, $\Delta\bar{\Delta}$&&
\\  \hline
$2^+2^{-+}$&$N_8^{\frac{3}{2}}\bar{\Delta}_8^{\frac{1}{2}}$,
$\Delta_8^{\frac{1}{2}}\bar{N}_8^{\frac{3}{2}}$,&939+1232&-7&-34
\\
& $N\bar{\Delta}$, $\Delta\bar{N}$, $\Delta\bar{\Delta}$&&
\\  \hline
$2^-3^{--}$&$\Delta\bar{\Delta}$&1232+1232&0&0 \\ \hline
$3^-0^{-+}$&$\Delta_8^{\frac{1}{2}}\bar{\Delta}_8^{\frac{1}{2}}$,
$\Delta\bar{\Delta}$&1232+1232&-88&-76\\ \hline
$3^+1^{--}$&$\Delta_8^{\frac{1}{2}}\bar{\Delta}_8^{\frac{1}{2}}$,
$\Delta\bar{\Delta}$&1232+1232&0&-67\\ \hline
$3^-2^{-+}$&$\Delta\bar{\Delta}$&1232+1232&0&0 \\ \hline
$3^+3^{--}$&$\Delta\bar{\Delta}$&1232+1232&0&0 \\ \hline \hline
\end{tabular}
\end{table}

The lowest multichannel coupling results for all
possible quantum numbers are listed in Table III, the superscript and
subscript of $N (\Delta)$ represent spin quantum number and color
dimensions, respectively. $E_T(B+\bar{B})$ is the threshold of decaying
into a baryon and an anti-baryon, $\Delta E_I$ and $\Delta E_{II}$ are
binding energies of hexaquark states $q^3\bar{q}^3$ in Model I
and Model II, respectively. It can be seen
from Table III that the states with $I^GJ^{PC}=0^-3^{--}$,
$1^+3^{--}$ and $3^+1^{--}$ are bound states only in Model II.
For other states almost the same qualitative results are obtained in two
models. It suggests that there are some bound states
below the lowest threshold in the present calculations.
The states with $I^GJ^{PC}=0^+0^{-+}$ and $1^-0^{-+}$ are
stable against disintegrating into $N$+ $\bar{N}$. The states with
$I^GJ^{PC}=1^-2^{-+}$, $2^-1^{--}$ and $2^+2^{-+}$ are stable
against disintegrating into $N$+$\bar{\Delta}$ or
$\bar{N}$+$\Delta$, but decay to $N\bar{N}\pi$ is allowed.
The states with $I^GJ^{PC}=0^+2^{-+}$,
$2^+0^{-+}$ and $3^+0^{-+}$ are stable against disintegrating into
$\Delta$+$\bar{\Delta}$, decaying to $N\bar{N}\pi\pi$ is allowed.
The states with $I^GJ^{PC}=0^-1^{--}$,
$1^+1^{--}$, $2^-3^{--}$, $3^-2^{-+}$ and $3^+3^{--}$ states are
not bound both in two models. The
multibody confinement potential based on the color flux tube
picture can give more attraction than the additive
two-body confinement interaction which is proportional to color
factors used in early multiquark state calculations, due to avoiding
the appearance of the anti-confinement in a color symmetrical quark or
antiquark pair. In fact one gluon exchange and one
boson exchange interaction also provide attractive interaction
for some states~\cite{hexaquarks}.

For the state $I^GJ^{PC}=0^+0^{-+}$, the wavefunction can be
separated into two groups, $N\bar{N}+N_8\bar{N}_8$ and
$\Delta\bar{\Delta}+\Delta_8\bar{\Delta}_8$. In Model I, there is no
interaction between $N\bar{N}+N_8\bar{N}_8$ and
$\Delta\bar{\Delta}+\Delta_8\bar{\Delta}_8$ due to the absence of
boson exchange term. The states $N\bar{N}+N_8\bar{N}_8$ and
$\Delta\bar{\Delta}+\Delta_8\bar{\Delta}_8$ have the lowest energies
1834 MeV and 2376 MeV, respectively. However,
$N\bar{N}+N_8\bar{N}_8$ and
$\Delta\bar{\Delta}+\Delta_8\bar{\Delta}_8$ are mixed in Model II
because there is interaction among them due to one boson exchange.
But the mixing effect is not large. The energies of
$N\bar{N}+N_8\bar{N}_8$ and
$\Delta\bar{\Delta}+\Delta_8\bar{\Delta}_8$ are 1865 MeV and 2384
MeV in the Model II if the mixing effect among two groups is
neglected. The mixing moves the energies of $N\bar{N}+N_8\bar{N}_8$
to 1844 MeV and $\Delta\bar{\Delta}+\Delta_8\bar{\Delta}_8$ to
2388 MeV. $N\bar{N}+N_8\bar{N}_8$ and
$\Delta\bar{\Delta}+\Delta_8\bar{\Delta}_8$ are bound states because
their energies are lower than the corresponding thresholds of
$N\bar{N}$ and $\Delta\bar{\Delta}$ in these two models. The masses
of $N\bar{N}+N_8\bar{N}_8$ and
$\Delta\bar{\Delta}+\Delta_8\bar{\Delta}_8$ states are close to the
masses of newly observed states $X(1835)$ and $X(2370)$, so it is
possible to interpret the main components of $X(1835)$ and $X(2370)$
as $N\bar{N}+N_8\bar{N}_8$ and
$\Delta\bar{\Delta}+\Delta_8\bar{\Delta}_8$ in the present
calculation, respectively. However, another state $X(2120)$ observed
by BES-III can not be described in the present calculations.

\begin{table}[ht]
\caption{Rms for $N\bar{N}+N_8\bar{N}_8$ and
$\Delta\bar{\Delta}+\Delta_8\bar{\Delta}_8$ (fm).}
\begin{tabular}{c|c|cccccccccc}\hline\hline
Model&Distances&$\left\langle
{\mathbf{R}_{qqq}}^2\right\rangle^{\frac{1}{2}}$&$\left\langle
{\mathbf{R}_{\bar{q}\bar{q}\bar{q}}}^2\right\rangle^{\frac{1}{2}}$&$\left\langle
\mathbf{X}^2\right\rangle^{\frac{1}{2}}$\\\hline
Model I &$N\bar{N}+N_8\bar{N}_8$&0.61&0.61&0.51\\
&$\Delta\bar{\Delta}+\Delta_8\bar{\Delta}_8$&0.65&0.65&0.60
\\\hline
Model II &$N\bar{N}+N_8\bar{N}_8$&0.66& 0.66&0.58\\
&$\Delta\bar{\Delta}+\Delta_8\bar{\Delta}_8$&0.71&0.71&0.66
\\\hline\hline
\end{tabular}
\end{table}

Using the wavefunctions of $N\bar{N}+N_8\bar{N}_8$ and
$\Delta\bar{\Delta}+\Delta_8\bar{\Delta}_8$, the root mean square
radii (rms) of the two states with $I^GJ^{PC}=0^+0^{-+}$ are
calculated and given in Table IV, where
\[\mathbf{R}_{qqq}=\mathbf{r}_i-\frac{\mathbf{r}_i+\mathbf{r}_j
+\mathbf{r}_k}{3}\]
and
\[\mathbf{R}_{\bar{q}\bar{q}\bar{q}}=\mathbf{r}_l-\frac{\mathbf{r}_l
+\mathbf{r}_m+\mathbf{r}_n}{3}.\]
It can be seen from Table IV that the radii are small and very close
in two models. The two clusters
$q^3$ and $\bar{q}^3$ are highly overlapped, therefore
the main components of $X(1835)$ and $X(2370)$ are not loose
hadronic molecule states but compact hexaquark states with
three-dimensional configurations similar to rugby ball in
the present calculations.

All hidden color components can not decay into two colorful hadrons
directly due to color confinement. $X(1835)$ and $X(2370)$ must
transform back into three color singlet mesons by means of breaking
and rejoining flux tubes before decaying into
$\eta^{\prime}\pi^+\pi^-$. This decay mechanism is similar to
compound nucleus formation and therefore should induce a resonance
which is named as a ``color confined, multiquark resonance"
state~\cite{resonance} in our models. It is different from all of
those microscopic resonances discussed by S.
Weinberg~\cite{weinberg}. Bicudo and Cardoso studied tetraquark
states using the triple flip-flop potential including two
meson-meson potentials and the tetraquark four-body potential. They
also found plausible the existence of resonances in which the
tetraquark component originated by a flip-flop potential is the
dominant one~\cite{Bicudo}.

\section{summary}

By using high precision few-body calculation method, GEM,
non-strange hexaquark states $q^3\bar{q}^3$ including $B_8\bar{B}_8$
and $B\bar{B}$ components are studied in flux tube models, extended
chiral quark model (Model II) and Isgur-Karl model (Model I), with a
six-body confinement potential. In the present version of flux tube
models, the system will automatically choose its favorable
configuration by means of the recombination of the flux tube when
the quarks and anti-quarks are moving. The flux tube models which
includes multibody confinement potential generally give more
attraction than the two-body confinement models with color factors
that was used in the early multiquark calculations. The two types of
flux tube models give similar results for non-strange hexaquark
system. Our calculations suggest that some states are stable against
decaying into a baryon and an anti-baryon. One gluon exchange and
one boson exchange interaction also provide attractive interaction
for some states, and therefore should be taken into account
altogether.

The states $X(1835)$ and $X(2370)$ can be explained as
$N\bar{N}+N_8\bar{N}_8$ and $\Delta\bar{\Delta}+\Delta_8\bar{\Delta}_8$
bound states in the flux tube models,
the main components are compact hexaquark states
$N_8\bar{N}_8$ and $\Delta_8\bar{\Delta}_8$, respectively. Such
states should be color confinement resonances with
three-dimensional configurations similar to rugby ball. $X(2120)$
can not be accommodated in this model. We admit that this
analysis is based on the mass calculation only, the decay properties of
these states have to be invoked to check the assignment, which is left for future.

\acknowledgments{This work is
supported partly by the National Science Foundation of China under
Contract Nos. 11047140, 11175088, 11035006, 11047023, and the Ph.D Program
Funds of Chongqing Jiaotong University.}

\end{document}